\documentclass{jetpl}
\twocolumn

\usepackage{amsmath,amssymb,epsfig,color,pstricks,bm,graphics,alltt}

\begin{document}

\lat

\title{Electronic Structure of New AFFeAs Prototype of Iron Arsenide 
Superconductors}

\rtitle{Electronic Structure of (Sr,Ca)FFeAs}

\sodtitle{Electronic Structure of New AFFeAs Prototype of Iron Arsenide Superconductors}

\author{I.\ A.\ Nekrasov$^+$, Z.\ V.\ Pchelkina$^*$, M.\ V.\ Sadovskii$^+$}

\rauthor{I.\ A.\ Nekrasov, Z.\ V.\ Pchelkina, M.\ V.\ Sadovskii}

\sodauthor{Nekrasov, Pchelkina, Sadovskii }

\sodauthor{Nekrasov, Pchelkina, Sadovskii }

\address{$^+$Institute for Electrophysics, Russian Academy of Sciences, 
Ural Division, 620016 Ekaterinburg, Russia}

\address{$^*$Institute for Metal Physics, Russian Academy of
Sciences, Ural Division, 620041 Ekaterinburg GSP-170, Russia}

\dates{Today}{*}

\abstract{
This work is provoked by recent discovery of new class prototype
systems AFFeAs (A=Sr,Ca) of novel layered ironpnictide High-T$_c$ superconductors
(T$_c$=36K). Here we report {\it ab initio} LDA results for electronic 
structure of the AFFeAs systems. We provide detailed comparison
between electronic properties of both new systems and reference LaOFeAs (La111) compound.
In the vicinity of the Fermi level all three systems have essentially the same
band dispersions. However for iron fluoride systems F(2$p$) states were
found to be separated in energy from As(4$p$) ones in contrast to La111,
where O(2$p$) states strongly overlaps with As(4$p$).
Thus it should be more plausible to include only Fe(3$d$) and As(4$p$) orbitals
into a realistic noninteracting model than for La111.
Moreover Sr substitution with smaller ionic radius Ca in AFFeAs materials
leads to a lattice contruction and stronger Fe(3$d$)-As(4$p$)
hybridization resulting in smaller value of the density of states
at the Fermi level in the case of Ca compound.
So to some extend Ca system reminds RE111 with later Rare Earths.
However Fermi surface of new fluorides is found to be nearly perfect two-dimensional.
Also we do not expect strong dependence of superconducting
properties with respect to different types of A substitutes.
}


\maketitle

At the beginning of 2008 new layered ironpnictide High-T$_c$
superconductors were found.
Up to now there are known several classes of such systems.
Following the order of discoveries:
1. Re111 (Re=La,Ce,Pr,Nd,Sm) with parent compound LaO$_{1-x}$F$_x$FeAs and T$_c$ about 40--55 K
\cite{kamihara_08,chen,zhu,mand,chen_3790,chen_3603,ren_4234,ren_4283};
2. A122 (A=Ba, Sr) with parent system Ba$_{1-x}$K$_x$Fe$_2$As$_2$ \cite{rott,ChenLi,Chu,Bud}
and T$_c$ about 38 K;
3. Li$_{1-x}$FeAs with T$_c$=18 K~\cite{cryst,wang_4688}.
Recently a new prototype material SrFFeAs was reported~\cite{Tegel,Han}.
This system has characteristic for ironarsenides SDW anomaly at about 175 K.
Later on this system doped with Co showed superconductivity at
T$_c \sim$5~K~\cite{Matsuishi}.
Finally for the system Sr$_{1-x}$La$_x$FFeAs superconductivity with
T$_c$=36~K was obtained by Zhu~$et~al.$~\cite{Zhu}. 
Also Zhu~$et~al.$~\cite{Zhu} provide some
crystallographic and resistivity data on CaFFeAs and EuFFeAs compounds.
Inspired by this new step of ironpnictides developments
we continue our work on these materials~\cite{Nekr,Nekr2,Nekr3} and propose
here first principle investigation of electronic
structure of fluoride compounds.

LDA (local density approximation) calculated electronic structure of La111 was addressed
in Refs.~\cite{singh,dolg,mazin} and is related with one for LaOFeP~\cite{lebegue}.
Investigation of RE111 series showed band structure to be insensitive to
the type of a Rare Earth~\cite{Nekr}. Comparative study of electronic properties
of other prototype systems BaFe$_2$As$_2$~\cite{Shein, Krell} and LiFeAs
is presented in Refs.~\cite{Nekr2,Nekr3}.

The manuscript is organized as follows: first we describe
crystal structure of (Sr,Ca)FFeAs compounds and computational details.
Then we provide circumstantial comparison between La111~\cite{Nekr} and
both Sr and Ca compounds with respect to the LDA 
computed band dispersions,
total, partial and orbitally resolved densities of states.
Short summary finalize our paper.

\section{Crystallographic and computational details }\label{str}

First crystal structure data for SrFFeAs were reported by Tegel~$et~al.$~\cite{Tegel}.
As well as other FeAs-systems~\cite{kamihara_08,rotter_4021,cryst}
SrFFeAs has tetragonal structure with the 
space group $P$4/$nmm$.
Lattice parameters are $a=3.9930(1)$~\AA~and $c=8.9546(1)$~\AA~\cite{Tegel}.
Atoms occupy following Wyckhoff positions~\cite{Tegel,param}:
F(2a) (0.75,0.25,0), Fe(2b) (0.75, 0.25, 0.5) and
Sr(2c) (0.25,0.25,0.1598), As(2c) (0.25,0.25,0.6527).
This crystal structure was confirmed by
several groups~\cite{Han,Matsuishi,Zhu}.
Also Zhu~$et~al.$~\cite{Zhu} determined lattice parameters
of CaFFeAs as $a=3.879$~\AA, $c=8.601$~\AA.
In Ref.~\cite{Zhu} no analysis of atomic positions for Ca system was given.
Thus for the time being in our calculations for Ca compound 
we use the same atomic positions as for Sr one.
General appearance of the Sr(Ca)FFeAs crystal structure
is similar to one of La111 and so we refer reader to the Refs.~\cite{Nekr,Nekr2}.

Main building blocks of new Sr and Ca systems
are AF and FeAs tetrahedra layers with combined cation
state ``+1'' and ``-1'' correspondingly.
To this end AF layer is equivalent to the
REO layer of RE111 systems~\cite{kamihara_08}.
However larger ionic radius of Sr in comparison with, for example, La,
gives sizeable expansion along $c$-axis of SrFFeAs structure, while
$a$ and $b$ are almost not affected.
Smaller ionic radii of Ca results in lattice contruction
of CaFFeAs relative to Sr one. Nevertheless $c$ lattice parameter
of CaFFeAs is of the same size as for
PrOFeAs compound~\cite{ren_4283}.
Physically important interatomic distances within
FeAs$_4$ tetrahedron layer Fe-Fe and Fe-As are practically
identical to those of La111.

Linearized muffin-tin orbitals method (LMTO)~\cite{LMTO}
was employed to calculate electronic structure of Sr(Ca)FFeAs
compounds.

\begin{figure}[!h]
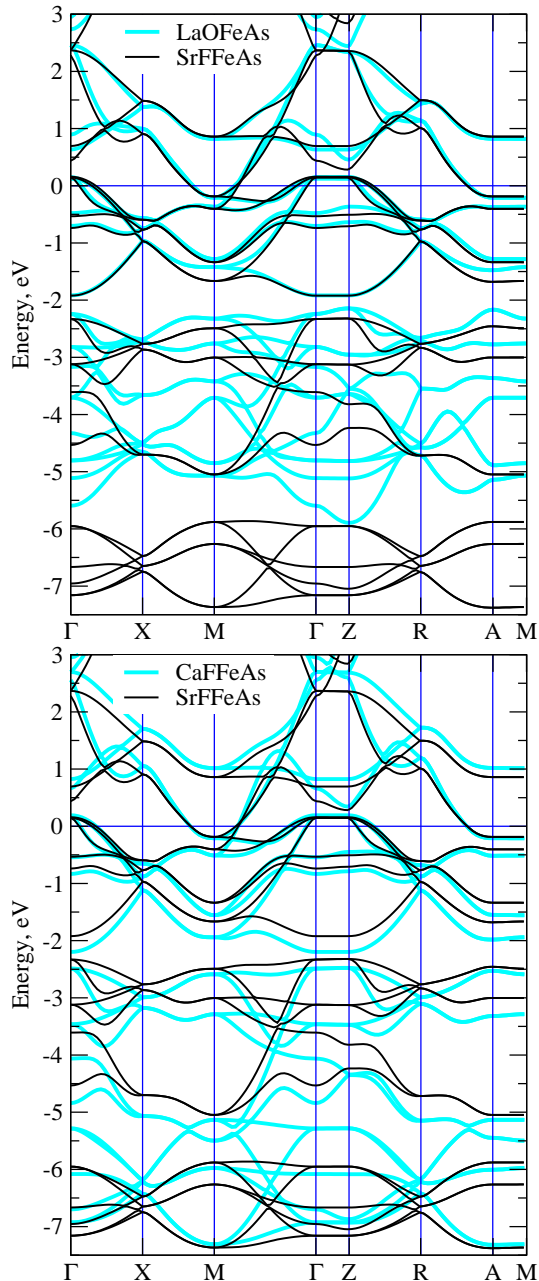

\includegraphics[clip=true,width=0.4\textwidth]{bands_LaSr.eps}
\includegraphics[clip=true,width=0.4\textwidth]{bands_SrCa.eps}
\label{bands}
\caption{Fig. 1. Comparison of LDA band dispersions
for La111 (wide gray lines) and SrFFeAs (black lines) on the upper panel
and for CaFFeAs (wide gray lines) and SrFFeAs (black lines) on the lower panel.
The Fermi level corresponds to zero.}
\end{figure}

\section{Results and discussion}\label{lda}
 
Upper panel of Fig. 1 shows comparison of band dispersions obtained within LDA
for new SrFFeAs (black lines) compound with one recently published for La111~\cite{Nekr}
(gray lines).
Bands crossing Fermi level are essentially identical for both compounds and belong
to Fe(3$d$) states (from -2~eV to 2.5~eV).
Main difference comes in the region of As(4$p$) states (-5.5~eV to -2~eV).
Here for La111 As(4$p$) and O(2$p$) states are essentially overlapped.
For SrFFeAs As(4$p$) states stay still in
interval (-5.5;-2)~eV but F(2$p$) states are completely separated from them
in energy and are located below between -7.5 and -6 eV.

In the lower panel of Fig.~1 we compare band dispersions
of Ca (gray lines) and Sr (black lines) iron fluorides.
Band structure of both systems reminds each other.
Again bands crossing Fermi level for Ca compound do not change much
from those of Sr one. However because of lattice contruction
and thus stronger hybridization effects
Fe(3$d$) band for CaFFeAs is slightly wider.
For the same reason As(4$p$) bands are slightly
lower in energy than for Sr material.
The F(2$p$) bands are most affected by both lattice
contruction and Sr to Ca substitution.
First one leads to about 1~eV broadening second one
to the energy elevation. So after all F(2$p$) bands somewhat intersect
with As(4$p$) bands.
One can note that to some extend difference between Sr and Ca iron fluorides
resembles very much difference between La111 and Pr111 iron arsenides (see~Ref.~\cite{Nekr})
although former one is larger and not that uniform.

Fig.~2 dispays direct matching of LDA calculated DOS
of Sr (solid black line) and Ca (dashed black line) iron fluorides with
La111 (gray line)~\cite{Nekr}. Upper panel contains
total DOS of all three systems.
The values of total DOS on the Fermi level are 3.65 state/eV/cell
for CaFFeAs, 4.27 state/eV/cell for SrFFeAs and 4.01  state/eV/cell
for reference La111 system, i.e. are only slightly different from each other.

In the energy interval of Fe(3$d$) states DOSes of all compounds
are rather repetitive (see also next from top panel of Fig.~2).
Moderate hybridization is observed between Fe(3$d$) and As(4$p$)
states for both fluorides and practically no hybridization 
between Fe(3$d$) and F(2$p$) as well as between As(4$p$) and F(2$p$) states 
(see next from bottom panel of Fig.~2).
For the Sr compound fluorine 2$p$ states are observed to be separated from 
all Fe(3$d$) and As(4$p$) states.
And for the case of Ca system F(2$p$) state touch As(4$p$) ones.

Finally we present orbitally resolved Fe(3$d$) DOS for Sr and Ca iron
arsenide-fluorides in Fig.~3. Bird eye picture looks like the
one for La111 (see Ref.~\cite{Nekr2}): Fermi level is crossed by
bands of predominantly $t_{2g}$ symmetry -- 
$xz$, $yz$, $x^2-y^2$ (later because of $\pi/4$ rotation of
local coordinate system around $c$-axis).
Nevertheless there are several fine distinctions with La111.
The $x^2-y^2$ orbital is strongly changed around 0.5~eV.
Some changes can be seen also for $3z^2-r^2$ orbital around
-0.5 eV and $xz$, $yz$ orbitals around -1.5~eV (see also Fig. 1, upper panel).
The same is valid also for Ca system except some bands broadening.
By looking on the Fig. 3 one can also understand
lowering of the total density of states for Ca system with respect to
Sr one mentioned above. Largest contribution to the DOS at the Fermi level comes from the
$x^2-y^2$ orbital for all of FeAs systems~\cite{Nekr,Nekr2,Nekr3}.
Because of stronger Fe-As hybridization it becomes wider and at the same time
lower in intensity to keep normalization.

In Fig. 4 the Fermi surface (FS) of the new Sr
fluoride system is compared with that of LaOFeAs~\cite{Nekr2}. 
Most important contrast to La111 FS is nearly perfect two-dimensional character
of the FS of SrFFeAs. All cylinders are practically ideal especially two large 
ones around $\Gamma$-point. FS for Ca fluoride system is practically
identical to that of Sr fluoride and we do not show it here.

\begin{figure}
\includegraphics[clip=true,width=0.9\columnwidth]{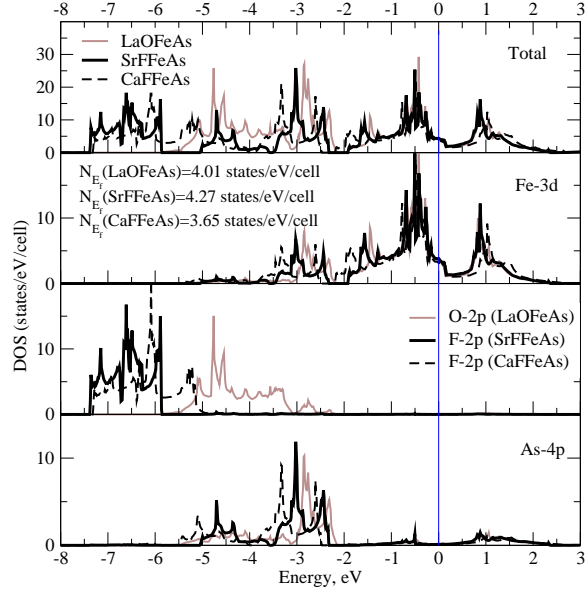}
\caption{\label{dos} 
Fig. 2. Comparison of LDA calculated total and partial DOS 
for SrFFeAs (solid black line), CaFFeAs (dashed black line)
and La111~\cite{Nekr} (gray line).
Panels from top to bottom: total DOS, Fe(3$d$), O(2$p$) and F(2$p$),
As(4$p$). The Fermi level corresponds to zero.}
\end{figure}

\begin{figure}
\includegraphics[clip=true,width=0.9\columnwidth]{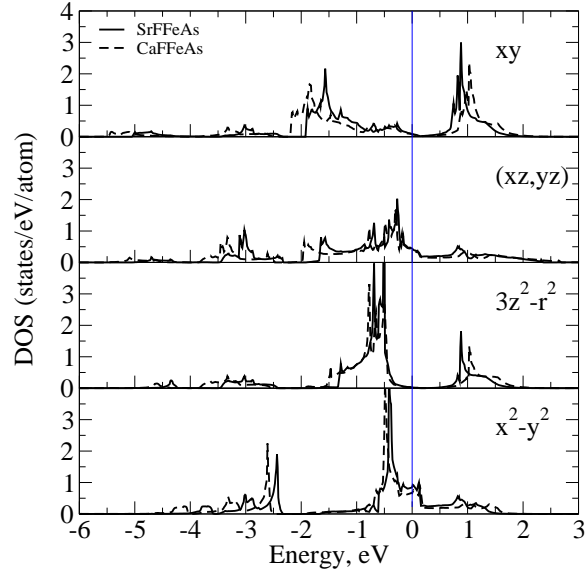}
\caption{\label{pdos}Fig. 3. Comparison of LDA orbitally resolved Fe-3$d$ DOS
for SrFFeAs (solid line) and CaFFeAs (dashed line).
The Fermi level corresponds to zero.}
\end{figure}

\begin{figure}
\includegraphics[clip=true,width=0.9\columnwidth]{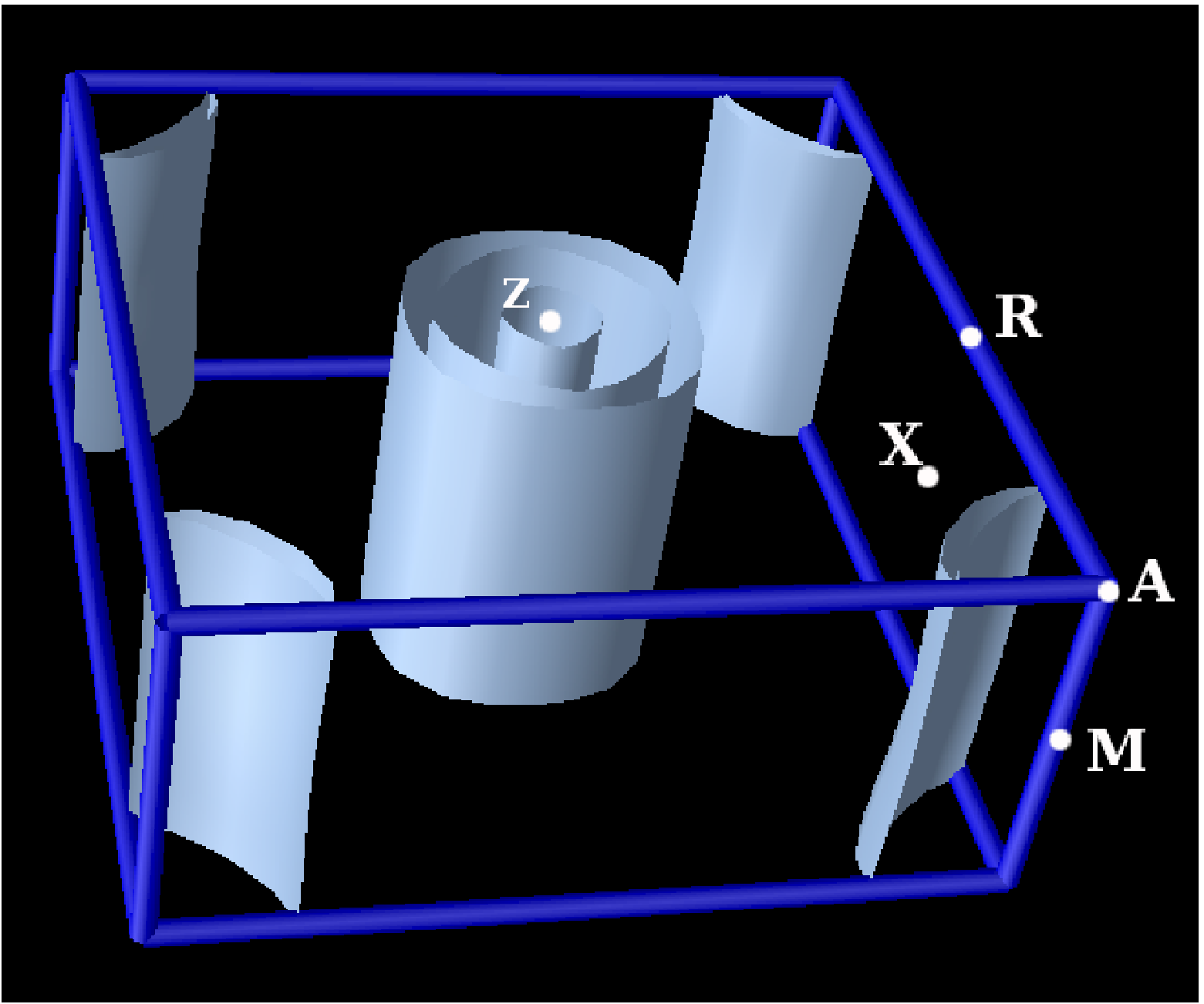}
\includegraphics[clip=true,width=0.9\columnwidth]{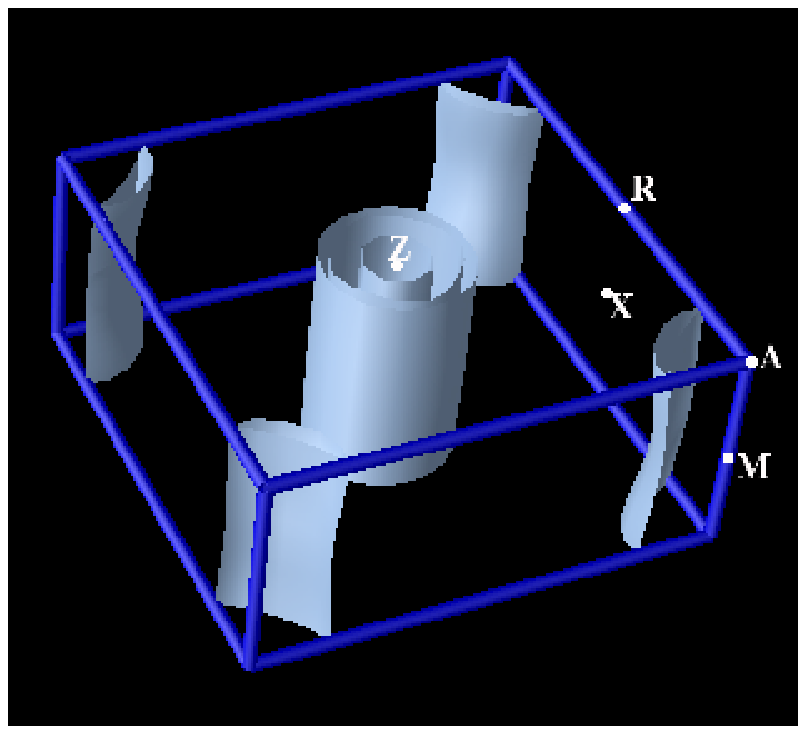}
\caption{\label{fs}Fig. 4. Comparison of LDA calculated Fermi surfaces
for SrFFeAs (upper panel) and LaOFeAs (lower panel).}
\end{figure}

\section{Conclusion}

LDA calculated electronic structure of new AFFeAs (A=Sr,Ca) prototype
of novel high-T$_c$ ironpnitides superconductors is presented here in
comparison with previously calculated~\cite{Nekr,Nekr2} for LaOFeAs material.
Despite different chemical contents new systems and La111 have identical crystal 
structures. Thus in fluoride systems electronic properties are also
determined by two-dimensional FeAs$_4$ layers as well.
However LDA obtained Fermi surface of new fruorides
is the most two-dimensional one among other known ironpnictides.
Main distinction of fluoride systems from La111
is energy separation between F(2$p$) state and
As(4$p$) states. Also practically no
hybridization of F(2$p$) and As(4$p$) or Fe(3$d$) states
is observed. It makes non interacting realistic model
of fluorides including only As(4$p$) and Fe(3$d$) states
well justified and easier to construct.
The CaFFeAs with contructed lattice (since Ca ion is smaller)
to some extend reminds RE111 systems with later RE.
In this work we do not observe any significant
difference of electronic structure of new fluorides and La111
close to the Fermi level,
thus we do not expect any remarkable changes of superconducting
properties with respect to different A substitutes. Lower values of 
DOS at the Fermi level for Ca compound probably makes it slightly less
promising in this respect.

\section{Acknowledgements}

This work is supported by RFBR grants 08-02-00021, 08-02-00712, RAS programs 
``Quantum macrophysics'' and ``Strongly correlated electrons in 
semiconductors, metals, superconductors and magnetic materials'',
Grants of President of Russia MK-2242.2007.2(IN), MK-3227.2008.2(ZP)
and scientific school grant SS-1929.2008.2,  interdisciplinary 
UB-SB RAS project, Dynasty Foundation (ZP) and Russian Science Support 
Foundation(IN). IN and MS thank University of Osaka (Japan) for 
hospitality during the visit in October 2008.


\begin{thebibliography}{99}

\bibitem{kamihara_08} Y. Kamihara, T. Watanabe, M. Hirano, H. Hosono, J. Am. Chem. Soc. {\bf 130}, 3296-3297 (2008).

\bibitem{chen}G.F. Chen, Z. Li, G. Zhou, D. Wu, J. Dong, W.Z. Hu, P. Zheng, 
Z.J. Chen, J.L. Luo, N.L. Wang, Phys. Rev. Lett. {\bf 101}, 057007 (2008).

\bibitem{zhu}X. Zhu, H. Yang, L. Fang, G. Mu, H.-H. Wen, Supercond. Sci. Technol. {\bf 21}, 105001 (2008).

\bibitem{mand}A.S. Sefat, M.A. McGuire, B.C. Sales, R. Jin, J.Y. Hove, 
D. Mandrus, Phys. Rev. B {\bf 77}, 174503 (2008).

\bibitem{chen_3790} G. F. Chen, Z. Li, D. Wu, G. Li, W. Z. Hu, J. Dong, P. Zheng, J. L. Luo, N. L. Wang,
Phys. Rev. Lett. {\bf 100}, 247002 (2008).

\bibitem{chen_3603} X. H. Chen, T. Wu, G. Wu, R. H. Liu, H. Chen, D. F. Fang, Nature {\bf 453}, 761 (2008).

\bibitem{ren_4234} Z.-A. Ren, J. Yang, W. Lu, W. Yi, X.-L Shen, G.-C. Che, 
L.-L. Sun, F. Zhou, Z.-X. Zhao, Europhys. Lett. {\bf 82}, 57002 (2008).

\bibitem{ren_4283} Z.-A. Ren, J. Yang, W. Lu, W. Yi, G.-C. Che, X.-Li. Dong, 
L.-L. Sun, Z.-X. Zhao, Materials Research Innovations {\bf 12}, 105 (2008).

\bibitem{rott}M. Rotter, M. Tegel, D. Johrendt, Phys. Rev. Lett. {\bf 101}, 107006 (2008).

\bibitem{ChenLi}G.F. Chen, Z. Li, G. Li, W.Z. Hu, J. Dong, X.D. Zhang,
N.L. Wang, J.L. Luo, Chin. Phys. Lett. {\bf 25}, 3403 (2008).

\bibitem{Chu}K. Sasmal, B. Lv, B. Lorenz, A. Guloy, F. Chen, Y. Xue, C.W. Chu,
Phys. Rev. Lett. {\bf 101}, 107007 (2008).

\bibitem{Bud}N. Ni, S.L. Bud'ko, A. Kreyssig, S. Nandi, G.E. Rustan, A.I. Goldman,
S. Gupta, J.D. Corbett, A. Kracher, P.C. Canfield, Phys. Rev. B {\bf 78}, 014507 (2008).

\bibitem{cryst} J.H. Tapp, Z. Tang, Bing Lv, K. Sasmal, B. Lorenz, Paul C.W. Chu, A, M. Guloy,
Phys. Rev. B {\bf 78}, 060505(R) (2008).

\bibitem{wang_4688}X.C.Wang, Q.Q. Liu, Y.X. Lv, W.B. Gao, L.X.Yang, R.C.Yu, F.Y.Li, C.Q. Jin,
arXiv: 0806.4688.


\bibitem{Tegel} M. Tegel, S. Johansson, V. Weiss, I. Schellenberg, W. Hermes, R. Poettgen, D. Johrendt,
arXiv:0810.2120v1.

\bibitem{Han}F. Han, X. Zhu, G. Mu, P. Cheng, H.H. Wen,
arXiv:0810.2475v1.

\bibitem{Matsuishi} S. Matsuishi, Y. Inoue, T. Nomura, M. Hirano, H. Hosono,
arXiv:0810.2351v1.

\bibitem{Zhu}X. Zhu, F. Han, P. Cheng, G. Mu, B. Shen, Hai-Hu Wen,
arXiv:0810.2531v2.

\bibitem{Nekr}I.A. Nekrasov, Z.V. Pchelkina, M.V. Sadovskii. Pis'ma Zh. Eksp.
Teor. Fiz. {\bf 87}, 647 (2008) [JETP Letters {\bf 87} (2008)],
arXiv: 0804.1239.

\bibitem{Nekr2}I.A. Nekrasov, Z.V. Pchelkina, M.V. Sadovskii, arXiv: arXiv:0806.2630,
JETP Letters, {\bf 88}, 144 (2008).

\bibitem{Nekr3}I.A. Nekrasov, Z.V. Pchelkina, M.V. Sadovskii, arXiv:0807.1010,
Pis'ma Zh. Eksp. Teor. Fiz. {\bf 88}, 621 (2008).

\bibitem{param} For our computations lattice constants of Sr and Ca
systems were taken from~Ref.~\cite{Zhu}.

\bibitem{singh} D.J. Singh, M.H. Du. Phys. Rev. Lett. {\bf 100}, 237003 (2008).

\bibitem{dolg}L. Boeri, O.V. Dolgov, A.A. Golubov, Phys. Rev. Lett. {\bf 101}, 026403 (2008).

\bibitem{mazin}I.I. Mazin, D.J. Singh, M.D. Johannes, M.H. Du,
Phys. Rev. Lett. {\bf 101}, 057003 (2008).

\bibitem{lebegue}S. Leb\`{e}gue, Phys. Rev. B {\bf 75}, 035110 (2007). 

\bibitem{Shein}I.R. Shein, A.L. Ivanovskii, arXiv: 0806.0750, 
Pis'ma Zh. Eksp. Teor. Fiz. {\bf 88}, 115 (2008).

\bibitem{Krell}C. Krellner, N. Caroca-Canales, A. Jesche, H. Rosner, A. Ormeci,
C. Geibel, Phys. Rev. B {\bf 78}, 100504(R) (2008).

\bibitem{rotter_4021} M. Rotter, M. Tegel, D. Johrendt, Phys. Rev. B {\bf 78}, 020503(R) (2008).

\bibitem{LMTO}O.K. Andersen. Phys. Rev. B {\bf 12} 3060 (1975);
O. Gunnarsson, O. Jepsen,  O.K. Andersen. Phys. Rev. B {\bf 27} 7144 (1983);
O.K. Andersen, O. Jepsen.  Phys. Rev. Lett. {\bf 53} 2571  (1984).




\end{thebibliography}
\end{document}